\documentclass{aa}  

\usepackage{graphicx}
\usepackage{xcolor}
\usepackage{txfonts}
\graphicspath{{figures/}} 
\usepackage{hyperref}
\begin{document}
   \title{NLTE analysis of the methylidyne radical (CH) molecular lines in metal-poor stellar atmospheres}
   \author{S. A. Popa
          \inst{1,2},
          R. Hoppe\inst{1}, M. Bergemann\inst{1}, C. J. Hansen\inst{2},   B. Plez\inst{3}, 
          T.C. Beers\inst{4}}
   \institute{Max-Planck-Institute for Astronomy, Königstuhl 17, 69117 Heidelberg, Germany
   \and
    Heidelberg University, Grabengasse 1, 69120 Heidelberg, Germany
    \and
Laboratoire Univers et Particules de Montpellier, Univ Montpellier, CNRS, Montpellier, France
    \and
Department of Physics and Astronomy and JINA Center for the Evolution of the Elements, University of Notre Dame, Notre Dame, IN 46556, USA    
             }
   \date{}


  \abstract
   {}
   {An analysis of the CH molecule in non-local thermodynamic equilibrium (NLTE) is performed for the physical conditions of cool stellar atmospheres typical of red giants ($\log g$ = 2.0, $T_{\rm eff}$ = 4500\,K) and the Sun. The aim of the present work is to explore whether the $G$-band of the CH molecule, which is commonly used in abundance diagnostics of Carbon-Enhanced Metal-Poor (CEMP) stars, is sensitive to NLTE effects.} 
   {Local thermodynamic equilibrium (LTE) and NLTE theoretical spectra are computed with the MULTI code. We use one-dimensional (1D) LTE hydrostatic MARCS model atmospheres with parameters representing eleven red giant stars with metallicities ranging from $\textnormal{[Fe/H]}= -4.0$ to $\textnormal{[Fe/H]} = 0.0$ and carbon-to-iron ratios $\textnormal{[C/Fe] = 0.0, +0.7, +1.5, and +3.0}$. The CH molecule model is represented by $1\,981$ energy levels, $18\,377$ radiative bound-bound transitions, and 932 photo-dissociation reactions. The rates due to transitions caused by collisions with free electrons and hydrogen atoms are computed using classical recipes.}
   {Our calculations suggest that NLTE effects in the statistical equilibrium of the CH molecule are significant and cannot be neglected for precision spectroscopic analysis of C abundances. The NLTE effects are mostly driven by radiative over-dissociation, owing to the very low dissociation threshold of the molecule and significant resonances in the photo-dissociation cross-sections. The NLTE effects in the $G$-band increase with decreasing metallicity. When comparing the C abundances determined from the CH $G$-band in LTE and in NLTE, we show that the C abundances are always under-estimated if LTE is assumed. The NLTE corrections to C abundance inferred from the CH feature range from $+0.04$ dex for the Sun to $+0.21$ dex for a red giant with metallicity $\textnormal{[Fe/H]}=-4.0$.}
   {Departures from the LTE assumption in the CH molecule are non-negligible and NLTE effects have to be taken into account in the diagnostic spectroscopy based on the CH lines. We show here that the NLTE effects in the optical CH lines are non-negligible for the Sun and red giant stars, but further calculations are warranted to investigate the effects in other regimes of stellar parameters.}

   \keywords{molecular data -- molecular processes -- radiative transfer -- stars:carbon -- stars:abundances -- stars:late-type}

   \titlerunning{NLTE analysis of CH}
   \authorrunning{Popa et al.}

   \maketitle
%

\section{Introduction}
In the last decades, there has been increasing interest in metal-poor stars in the Milky Way (MW) and its dwarf satellite galaxies (see \citealt{BeersChrislieb} for a historical summary of the field at the time, and \citealt{Frebel_2018} for a more recent review). Being the oldest stars that can be directly observed at present, they carry invaluable information about the early Galaxy, including initial conditions for star and galaxy formation \citep[e.g.][]{klessen,frebel&norris,Yoon_2018}, large galaxy assembly \citep[e.g.][]{Carollo_2007,Carollo_2010, Beers_2012}, and the early history of nucleosynthesis \citep[e.g.][]{Hansen2011,gilpons_2021,Qian}.

As more and more metal-poor stars are observed, it has become clear that the frequency of stars that exhibit strong carbon enhancement (resulting in $[$C/Fe$]= +0.7$ or even higher) dramatically increases with decreasing metallicity. \citet{BeersChrislieb} classified such stars as Carbon-Enhanced Metal-Poor (CEMP) stars (using the criteria at the time of [Fe/H] $< -1.0$ and [C/Fe] $> +1.0$. The nature of the CEMP stars is reflected in their absolute carbon abundance ($A$(C); \citealt{spite2013,Yoon_2016}), as well as in their heavy-element content \citep[e.g.][]{BeersChrislieb,Frebel_2018,Hansen_2019}. 

The nature and chemical composition of the CEMP stars is not yet well-settled, and has been targeted in numerous observational studies \citep[e.g.][]{Aoki_2008,Aoki_2022, 
Yong_2013, Ito_2013, Lee_2013, Masseron, Yoon_2016, Rasmussen_2020, Zepeda_2022}. One sub-class of CEMP stars exhibits very high levels of C combined with a strong enrichment in slow neutron-capture (s-) process elements, and can be explained by mass transfer in a binary system from an asymptotic giant branch (AGB) star onto a smaller less evolved binary companion that we now observe as a CEMP-$s$ star \citep[e.g.][]{Aoki2007, Placco_2013, starkenburg, Abate2015, hansen2016s, Hansen_2019, caffau_2019, goswami}. Another sub-class, the CEMP-no stars, shows absolute C abundances that are 1-3 dex lower than the CEMP-$s$ stars, with no enhancements in neutron-capture elements. Although their origin is still debated, they are believed to be bona fide second-generation stars \citep[e.g.][]{Ryan_2005, hansen_2016, Zepeda_2022b}, where possible formation sites are polluted by massive fast-rotating ultra metal-poor (UMP; [Fe/H] $ < -4.0$) and hyper metal-poor (HMP; [Fe/H] $< -5.0$) stars, also known as spinstars \citep{Meynet2006,choplin,choplin_2020,liu} or faint core-collapse supernovae that undergo mixing and fallback \citep[e.g.][]{Nomoto2013},leaving behind large amounts of lighter elements like C but lower amounts of the heavy elements. This evidence makes CEMP-no stars a candidate for the first low-mass stars to form in the early universe \citep{hansen_2016,CJHansen2016}.

Owing to the low metallicities of CEMP stars, the main diagnostic of the C abundance is the $G$-band, which is the most prominent feature of the CH radical in their optical spectra  \citep{Masseron}. The $G$-band is found around 4300 {\AA} and it is also used as a diagnostic for the Morgan-Keenan (MK) classification scheme \citep{MK}. It consists of numerous molecular lines that arise from electronic transitions between the electronic ground state X\textsuperscript{2}$\Pi$ and the first excited electronic state A$\textsuperscript{2}\Delta$ of the CH radical.

Past analyses of the CH $G$-band were carried out under the assumption of Local Thermodynamic Equilibrium (LTE) \citep[e.g.][]{carbon2,carbon4,carbon3}, although it was shown early on that CH lines were formed out of LTE in the solar atmosphere \citep{EPP1960}.
The LTE assumption greatly simplifies the complexity of the computations. However, it also fails to realistically describe the interaction between radiation and matter in stellar atmospheres \citep{bergemannnordlander}. This is instead achieved by non-LTE (NLTE) modelling, where the radiation field and the distribution of matter over different energy states are not simply given by their equilibrium values, as in LTE, but are explicitly determined by solving the coupled radiative transfer and statistical equilibrium equations. 

This work presents the analysis of the NLTE statistical equilibrium for the CH radical. Our aim is to determine whether departures from LTE are significant, in particular for the conditions pertaining to cool, metal-poor stellar atmospheres. 

The paper is organised as follows. The stellar atmospheres and the molecule model are presented in Sect. \ref{sec:methods}. The modelling of theoretical spectra of the CH molecule is presented in Sect. \ref{sec:results}, along with the NLTE abundance corrections for representative model atmospheres for cool red giants of various metallicities and carbon enhancements. The results are discussed in the context of other works in Sect. \ref{sec:discussion}, and are summarised in Sect. \ref{sec:conclusion}. 
%
%
%
%
\section{Methods}
\label{sec:methods}
Due to the scarcity of literature on the NLTE modelling of molecules, we briefly recap the nomenclature. Electronic levels of molecular species are described by the orbital angular momentum $\Lambda$ ($\Sigma$, $\Pi$, and $\Delta$ for 0,1,2 states, accordingly), vibrational quantum number v, rotational quantum number $J$, the total electron angular momentum $\Omega$ (sum of $\Lambda$ and $\Sigma$, the electronic spin), and symmetry describing geometric operations (e$/$f). Following the standard notation in chemistry, we furthermore denote the ground state with a $'$ and an excited state with a $''$.
\subsection{Model of the CH radical}
\label{sec:ch_molecule}

\begin{figure}
\centering
\includegraphics[width=\columnwidth]{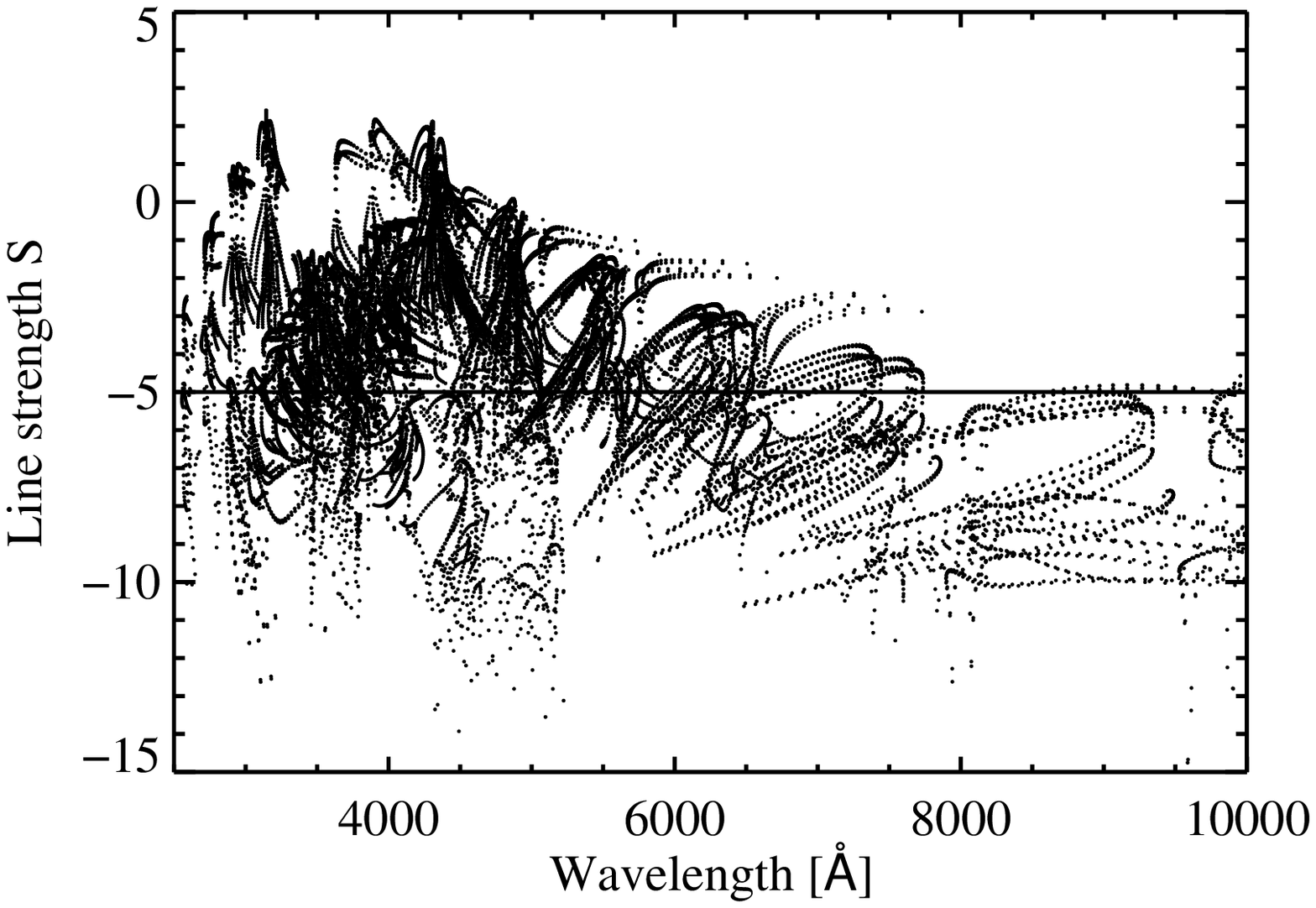}
\includegraphics[width=\columnwidth]{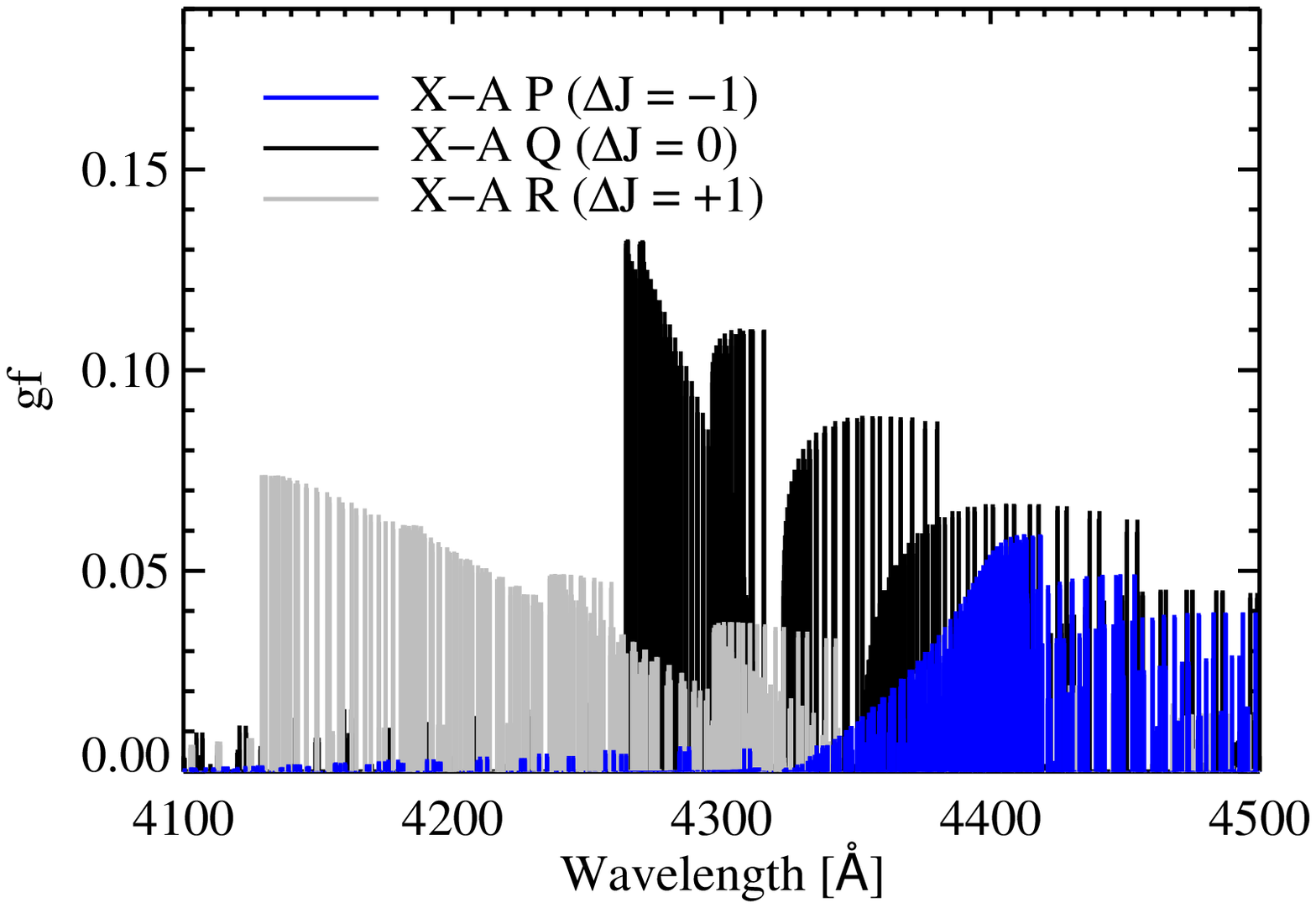}
\caption{Top panel: Distribution of line strengths in the entire CH system. Only transitions with strengths greater than $S=-5$ are included in the NLTE model molecule. Bottom panel: The X-A spectrum of the CH molecule around the wavelength of $\sim 4300$ \AA. The P-, Q-, and R-branches are indicated. The strongest feature in this range, corresponding to $J' - J" = 0$ (vibrational transition accompanied by a rotational transition) is the astrophysically famous '$G$-band'.}
\label{fig:ls}
\end{figure}
\begin{figure}[t]
    \centering
    \includegraphics{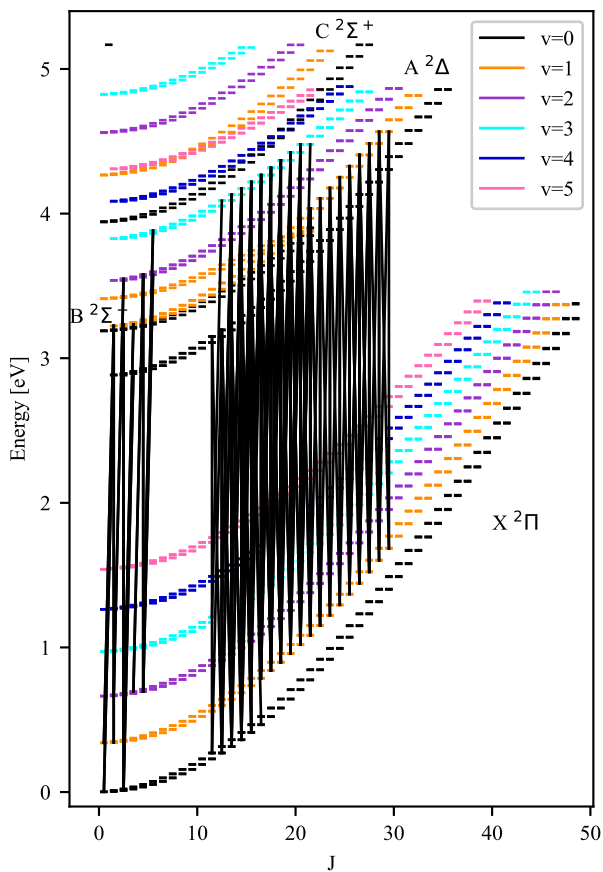}
    \caption{Energy levels included in the CH molecule model with specified rotational quantum number $J$ and vibrational quantum number $v$. The black lines indicate the bound-bound transitions between the electronic states A$^2\Delta$ and X$^2\Pi$ that give rise to the molecular $G$-band lines in wavelength range 4297 - 4303 {\AA}.}
    \label{fig:energy_levels}
\end{figure}
\begin{figure}
    \centering
    \includegraphics{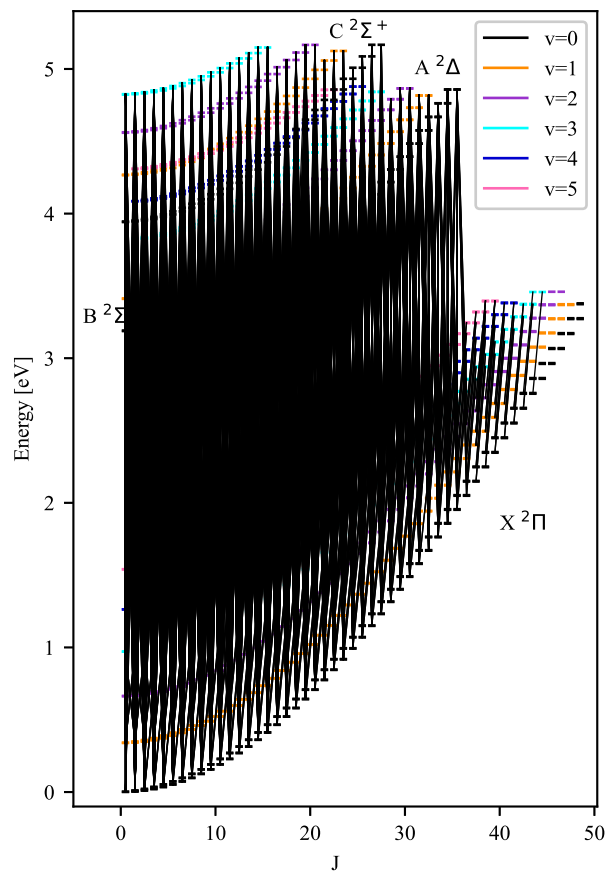}
    \caption{Energy levels included in the CH molecule model with specified rotational quantum number $J$ and vibrational quantum number $v$. The black lines indicate all radiative bound-bound transitions included in the model.}
    \label{fig:transitions}
\end{figure}

\begin{figure}
    \centering
    \includegraphics{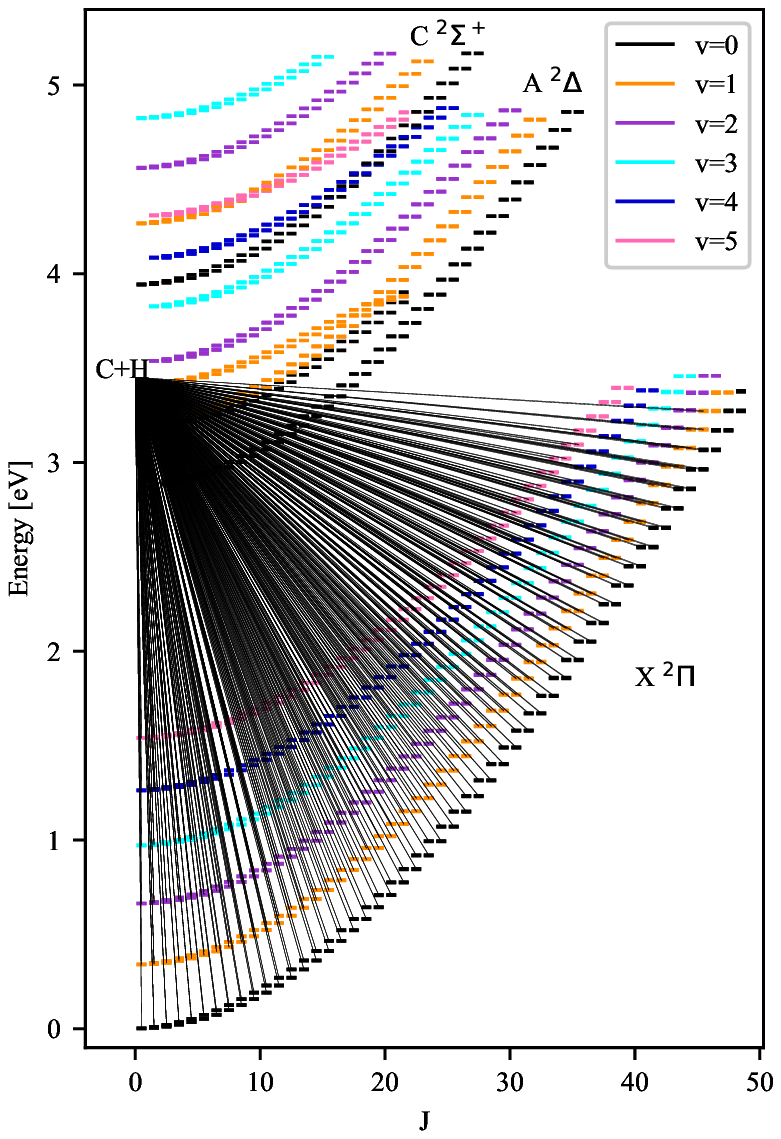}
    \caption{The black lines indicate the transitions for the photo-dissociation reactions with the dissociation energy of 3.45 eV.}
    \label{fig:transitions_photodis}
\end{figure}
CH is a heteronuclear diatomic molecule with an internuclear distance of 0.16 nm \citep{Demtroeder} and the binding energy of the ground state $^2\Pi$ of 3.45 eV \citep{Herzberg1969}. In our representation, the CH model contains $1\,981$ energy levels, $18\,377$ radiative bound-bound transitions, and $932$ photo-dissociation transitions. The data for energy states and bound-bound radiative transitions were taken from \citet{Masseron}. The latter model does not include the a$^4 \Sigma^{-}$ state, as there are no bound-bound radiative transitions from this to other states of the molecule, therefore this state is also not included in our model. The photo-dissociation cross-sections for the available states are adopted from \citet{photodisskurucz}. In our model, we retain all energy states with distinct molecular quantum numbers.  Whereas the original datasets comprise $52\,201$ radiative transitions, to simplify the calculations we retain only transitions in the wavelength range from $1\,000$ to $20\,000$ \AA, as this covers the significant range of the spectral energy distribution of cool stars, and astrophysical line strengths larger than $-5$ (Fig. \ref{fig:ls}, top panel) computed using the following expression:
\begin{equation}
    S = -(5039.44 \times E_{\rm low}/T) + log_{10}(\lambda \times gf),
\end{equation}
where the energy of the lower level $E_{\rm low}$ is given in eV, wavelength $\lambda$ in \AA, and $gf$ is the product of the oscillator strength $f$ and the statistical weight of the lower level $g$, and T the temperature in Kelvin, here set to $T=4000$ K. We also exclude the extremely weak transitions with f-values less than $10^{-7}$. Most of them connect energy states with small energy separations, and this may lead to numerical problems with population inversions in SE calculations. In Fig. \ref{fig:ls} (bottom panel), we also show the distribution of gf-values of the X-A transitions in the wavelength range around $\sim 4300$ \AA, which is easily accessible by astronomical facilities. The P-, Q-, and R-branches are indicated. The strongest feature in this range, corresponding to $J' - J" = 0$ (vibrational transition accompanied by a rotational transition) is the astrophysically famous '$G$-band'.

The energy levels included in the NLTE model are displayed in Figs. \ref{fig:energy_levels}, \ref{fig:transitions}, and \ref{fig:transitions_photodis} in the plane of J (rotational number) against level energy (in eV). Figure \ref{fig:energy_levels} shows the $G$-band lines arising in the transitions in the limited wavelength range from 4297 to 4340 \AA, Fig. \ref{fig:transitions} shows the complete set of radiative bound-bound transitions included in the model and Fig. \ref{fig:transitions_photodis} shows the photo-dissociation reactions. In our model, the photo-dissociation transitions take place between the ground state levels X$^2\Pi$ with energies $E=0-3.17$ eV, rotational quantum numbers $J=0.5-45.5$, and vibrational quantum numbers $v=0-5$ and the dissociated state C + H. As evident from the plots, the electronic ground state X$^2\Pi$ is clearly separated from the other excited electronic states A$^2\Delta$, B$^2\Sigma^-$ and C$^2\Sigma^+$, which have an intertwined structure in this particular representation, where the energy of the levels is plotted against the rotational number $J$. These diagrams reveal the complexity of the energetic structure of molecules, with each electronic energy level consisting of finer vibrational energy levels and even finer rotational energy levels. 

All bound energy levels are also connected through collisional transitions, where collisions with free electrons, as well as with hydrogen atoms, are considered. The collisional recipes assumed in our study are approximate \citep[see e.g.][]{Barklem2016}, however, for the lack of more accurate datasets, they provide a suitable starting point for exploring the effects of NLTE in the molecule. The rates of transitions caused by collisions with free electrons e$-$ are computed using van Regemorter's formula for allowed transitions \citep{Regemorter} and the Allen's formula \citep{Allen} for forbidden transitions ($\log gf \leq -10$). The rate processes in bound-bound, as well as bound-free transitions, caused by collisions with H atoms are computed using Drawin's formula in the formulation of \citep{Steenbock}. Limited theoretical studies of inelastic collisions of CH rotational/fine-structure levels with molecular hydrogen exist \citep{Dagdigian2017}, however, the temperatures considered (up to $300$ K) are tailored to the conditions of the ISM, and are too low for the purposes of our study. Extrapolation of these datasets to stellar conditions is not advised (P. Dagdigian, priv. comm.). The results of NLTE calculations in CH might be sensitive to collisions. However, a more detailed study of this aspect would require direct quantum-mechanical estimates of rate coefficients at least for a subset of relevant transitions (especially, for the $C+H$ reactions) and we defer such an analysis for future work, once such estimates become available for the conditions of atmospheres of FGK-type stars. 
\subsection{Statistical equilibrium calculations}
The calculations of NLTE statistical equilibrium and $G$-band spectra are done with the help of the MULTI code \citep*{ScharmerCarlsson, multi}. This code was developed in order to solve NLTE radiative transfer problems in semi-infinite, one-dimensional (1D) atmospheres with a prescribed macroscopic velocity field, where the model atom can have many atomic levels and several ionisation stages. The code uses an Approximate Lambda Iteration (ALI) method for the statistical equilibrium calculations, and the local version of the Lambda operator is used in this work. All calculations are carried out taking line blanketing into account. Detailed opacities were computed as described in \citet{Bergemann2019} using the MARCS model atmospheres and Turbospectrum code \citep{Plez2012,Gerber2022} for a range of metal mixtures representative of the model atmospheres studied in this work. The linelists used for the opacity calculations include all chemical elements up to U in the first 3 ionisation stages \citep[for more details, see][]{Bergemann2013,Bergemann2015}.

\begin{figure}
    \centering
    \includegraphics[width=\columnwidth]{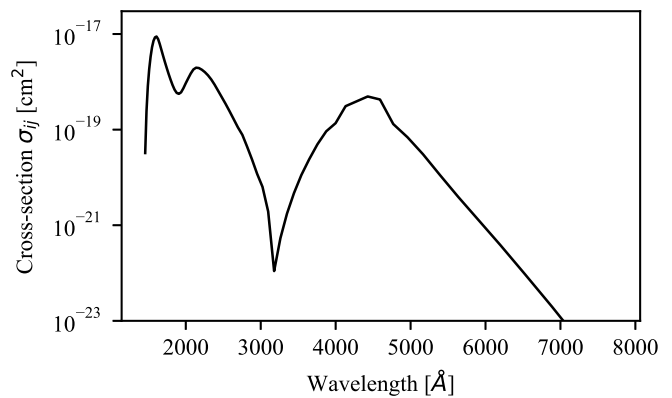}
    \caption{Cross-section distribution of the photo-dissociation for the energy level at 2.45 eV (19777.31 cm$^{-1}$) with the following configuration: electronic state X$^2\Pi$  with quantum numbers $v=0$ and $J=40$.  }
    \label{fig:photodis}
\end{figure}

For the CH radical, transitions from the bound levels to the continuum level correspond to photo-dissociation processes, which represent the breaking of the CH molecule into its constituent atoms, C and H, as described by the reaction:
\begin{equation}
    \textnormal{CH} + \gamma \rightarrow \textnormal{C}+ \textnormal{H}\,.
\end{equation}
The necessary energy for this reaction to take place from the ground state is given by the threshold dissociation energy of $3.45$ eV, which corresponds to the energy of $\sim 1.6$ eV below the last bound energy state of the molecule. The cross-sections for one of the energy states of CH is shown in Fig. \ref{fig:photodis}.
\begin{figure*}[t]
    \centering
    \includegraphics[width=\textwidth]{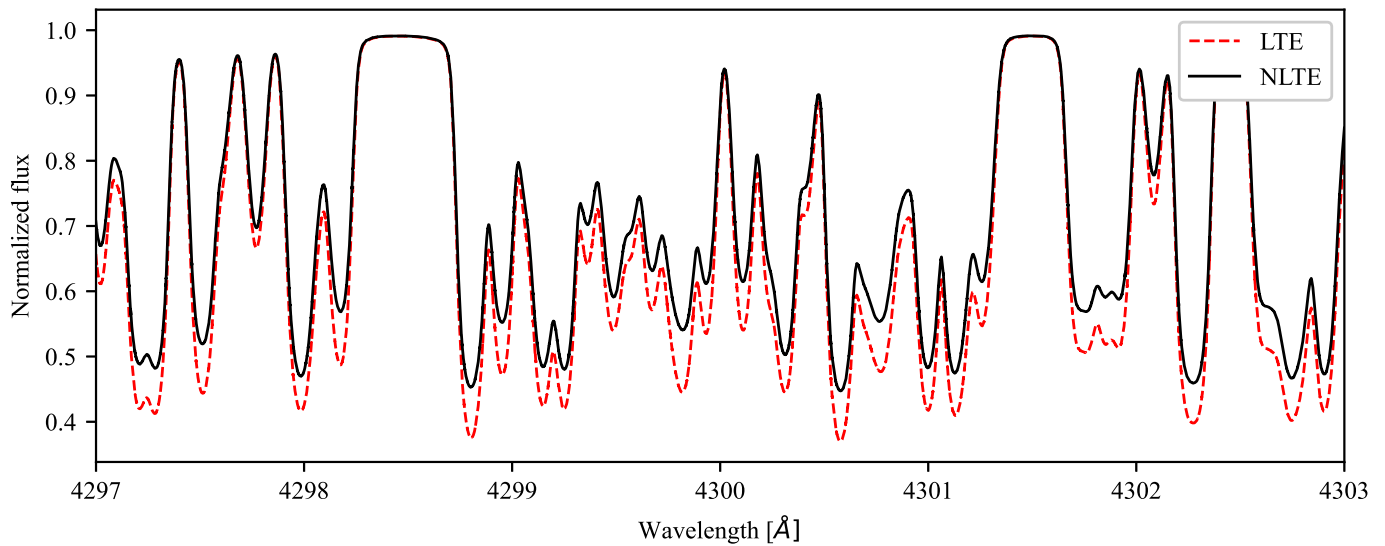}
    \caption{LTE and NLTE spectra in the region of the CH $G$-band, computed using the model atmosphere with a solar carbon abundance of $A$(C)=8.43.}
    \label{fig:gband_sunmarcs}
\end{figure*}
To compute the photo-dissociation rate coefficient $R_{ij}$, we follow the standard description of bound-free processes \citep[e.g.][]{HM2014}:
\begin{equation}
    n_i R_{ij} = n_i 4 \pi \int_{\nu_0}^\infty{ \frac{\sigma_{ij}(\nu)}{h\nu} J_\nu d\nu },
\end{equation}
with $n_i$ the number density of the CH level ($i$) that dissociates, $j$ the level in the C\,I atom to which it dissociates (all dissociations in our model molecule end with the neutral C atom in its ground state), $\sigma_{ij}(\nu)$ the photo-dissociation cross-section, $J_\nu$ the mean intensity of radiation, $\nu_0$ the threshold energy, and $\nu$ the frequency. The reverse reaction, namely the photo-attachment, which describes the formation of the CH molecule from individual C and H atoms, is computed assuming micro-reversibility as follows:
\begin{equation}
    n_j R_{ji} = n_j \left( \frac{n_i}{n_j}\right)^* 4\pi \int_{\nu_0}^\infty{\frac{\sigma_{ij}(\nu)} {h\nu} \left( \frac{2 h \nu^3}{c^2} + J_\nu\right) e^{-h\nu/kT} d\nu},
\end{equation}
where the star * symbol denotes LTE populations. Photo-attachment reactions can be associated with stimulated and spontaneous processes, and in this equation the factor $J$ represents the induced photo-attachment and the factor $2h \nu^3/c^2$ describes  spontaneous transitions. The ratio $(n_i^*/n_j^*)$ represents the ratio between the number density of CH molecules in the energy state $i$ relative to the number density of C I atoms in the ground state, and it is computed as:
\begin{equation}
  \frac{n_i^*}{n_j^*} = \left(n_{\rm CH}/n_{\rm C}\right) g_{i} e^{-E_{i}/kT}/Q_{\rm CH}/ (g_{0}/Q_{\rm C}),
\end{equation}
where $n_x$ and $Q_x$ are the total number densities and partition functions of the species $x$, $g_0$ is  the multiplicity of the C\,I ground state, and $g_i$ and $E_i$ are the multiplicity and energy of a given CH level $i$. The LTE C\,I level populations are computed using the Saha and Boltzmann equations. The CH number densities are calculated using the equilibrium constants from \citet{Tsuji1973}.

Photoionisation of CH is not important for the SE of the molecule, because of the very high ionisation potential of CH ground state ($10.6$ eV). This situation is analogous to O I, for which photo-ionisation processes and the details of the corresponding cross-sections have a negligible effect on the statistical equilibrium of the ion \citep{Bergemann2021,Bautista2022}. 
%
%
%
%
\subsection{Model atmospheres}
\label{sec:atmos}
In this work, we use the MARCS model atmospheres \citep{Gustafsson}. The models represent giant K-type stars ($\log g$ = 2.0, $T_{\rm eff}$ = 4500\,K) with metallicities in the range $-4.0 \leq {\rm [Fe/H]} \leq 0.0$ and microturbulence of 2 kms$^{-1}$. They are computed under the assumption of LTE and hydrostatic equilibrium, using a 1D geometry with a spherical symmetry. We note that in this paper we limit the analysis to K giants, because these types of stars are commonly used for spectral diagnostics of carbon abundances at very low metallicity. Other spectral types and luminosity classes will be investigated in our upcoming papers.

Because of the assumptions of 1D hydrostatic equilibrium, these model atmospheres do not describe time-dependent three-dimensional 
phenomena like convection and turbulence, hence, additional parameters have to be introduced. In the models, convection is described using the mixing-length theory presented by \citep{Henyey}, parametrised by \makebox{$\alpha = l/H_p = 1.5$}, where $l$ is the mixing length and $H_p = \frac{P}{\rho g}$ is the pressure scale-height. Another parameter is the microturbulent velocity, $\xi_t$, which describes turbulent motions on scales smaller than the photon's mean free path in the atmosphere. This parameter is taken to be 2 km/s.
\section{Results} \label{sec:results}

Molecular spectra are more complex than atomic spectra, due to their more complicated energy structure given by the different electronic, vibrational, and rotational energy levels, and the numerous molecular transitions that can take place between them. These transitions give rise to numerous absorption lines, which by overlapping create the absorption bands that are characteristic of molecular spectra. 

In the following, we analyse the properties of $G$-band spectra calculated in LTE and NLTE for the Sun and for cool red giant stars with different metallicities. All transitions that give rise to molecular lines in this range are displayed in Fig. \ref{fig:energy_levels}. 
\subsection{Sun}

The $G$-band in the spectrum of the model atmosphere with the parameters of the Sun ($T_{\rm eff}=5777$ K, 
$\log g = 4.44$, [Fe/H]$= 0.0$) is displayed in Fig. \ref{fig:gband_sunmarcs}. The wavelength range between \makebox{4297 - 4303 {\AA}} was chosen since it allows us to both visually distinguish the line profiles and interpret the effects of NLTE. As the $G$-band is formed from the transitions between the same electronic energy states, any chosen wavelength region would lead to the same results. Both LTE and NLTE profiles are computed using a C abundance of $A$(C) = 8.43  \citep{Asplund_sun}. Even though the same atmosphere and molecule model were used for the computation, the figure shows that the NLTE profile of the $G$-band is weaker than the LTE profile across the entire considered wavelength range. 
\begin{figure}
\includegraphics{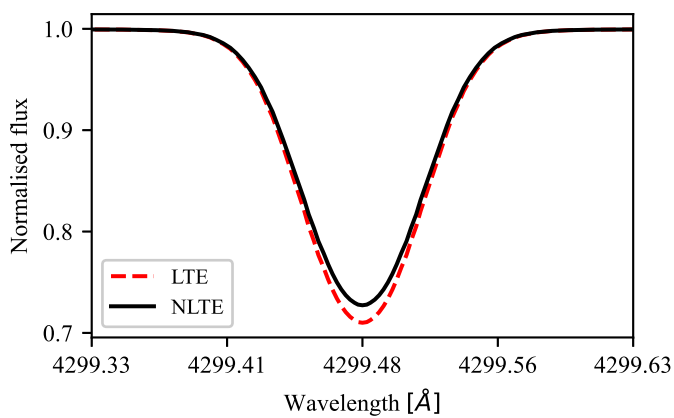}
\caption{The LTE and NLTE line profiles of the representative $G$-band line at 4299.48 {\AA}. }
\label{fig:line_0}
\end{figure}
\begin{figure}
\includegraphics{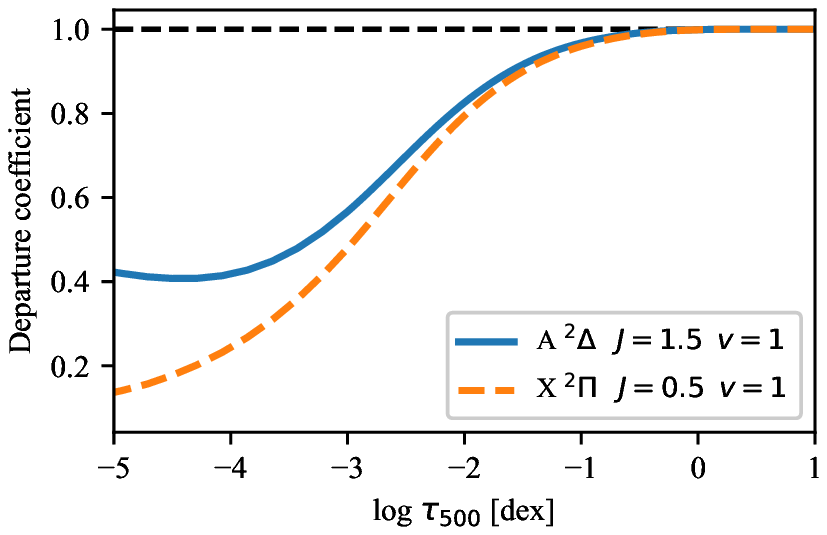}\\
\includegraphics{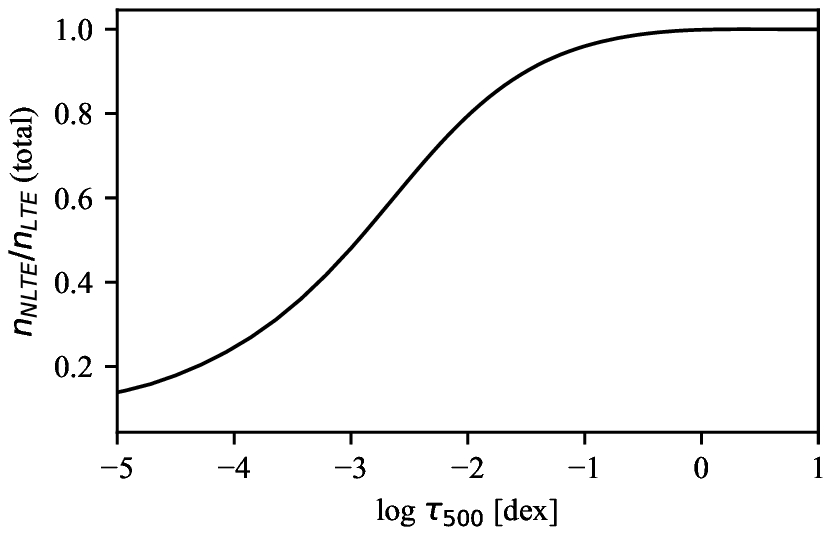}
\caption{Top panel: The departure coefficients for the upper level of the transition $A^2\Delta$ with $E=3.2$ eV, $J=1.5$ and $v=1$ and the lower level $X^2\Pi$ with $E=0.3$ eV, $J=0.5$ and $v=1$. The dashed black line indicates the departure coefficient for a thermalised level, for which $n_i^\textnormal{LTE}=n_i^\textnormal{NLTE}$ holds true. Bottom panel: Ratio of the total CH number density in NLTE to the total CH number density in LTE.}
\label{fig:dep_0}
\end{figure}
The difference between the NLTE and LTE line profiles is caused by the differences in the distribution of particles over excitation stages, and to the dissociation of CH. Therefore, it is useful to make use of the concept of the departure coefficient $b_i$ for a level $i$ \citep[e.g.][]{bergemannnordlander}:
\begin{equation}
    b_i=\frac{n_i^\textnormal{NLTE}}{n_i^\textnormal{LTE}}\,,
    \label{eq:b_i}
\end{equation}
where $n_i^\textnormal{LTE}$ is the population given by the equilibrium Saha-Boltzmann distribution, and $n_i^\textnormal{NLTE}$ is the actual population computed iteratively using the statistical equilibrium equations. If $b_i=1$, the population is given by the LTE value and the level is said to be thermalised, whereas $b_i\neq1$ indicates departure from LTE. If $b_i>1$, the level is over-populated, and if $b_i<1$ the level is under-populated with respect to the population given by the LTE assumption. 

To illustrate the difference between the LTE and NLTE computation, we inspect a representative $G$-band line at 4299.48 {\AA} (Fig. \ref{fig:line_0}). It arises from the transition between the electronic state $A^2\Delta$ with energy $E_j=3.2$ eV, rotational quantum number $J_j=1.5$ and vibrational quantum number $v_j=1$ and the ground state $X^2\Pi$ with $E_i=0.3$ eV, $J_i=0.5$ and $v_i=1$, respectively. The departure coefficients for the upper level $j$ and lower level $i$ of the representative line are displayed in Fig. \ref{fig:dep_0}, top panel. In the deeper layers of the atmosphere ($\log\tau_{500}\gtrsim0$), both levels are thermalised, $b_i=b_j=1$. However, in the upper photospheric layers ($\log\tau_{500}\lesssim0$), the populations start to experience departures from the LTE assumption. This corresponds to our expectations, as the LTE assumption is known to only hold in the deeper layers of the stellar atmosphere, where the density is high and collisions are sufficiently frequent to thermalise the atmosphere. The radiation field and the level populations are then given by their equilibrium distributions, which only depend on local quantities such as temperature and density. 

However, in the upper layers, the density and temperature decrease, photons travel long distances without coupling to local quantities, and radiative transitions outweigh collisional ones. This establishes a non-linear, non-local relation between the radiation field and the level populations, which is taken into account by the NLTE calculations. In the case of the considered spectral line at $4299.48$ \AA, the NLTE effects in the upper layers of the atmosphere cause both levels to be under-populated with respect to the LTE level populations, as displayed in Fig. \ref{fig:dep_0}, top panel. This NLTE effect is primarily driven by the dissociation of the CH molecules, owing to the low photo-dissociation energy threshold. This can be referred to as over-dissociation (analogous to the over-ionisation in atoms) and is displayed in Fig. \ref{fig:dep_0}, bottom panel. In the upper photospheric layers, the number density of CH in the NLTE computation gets increasingly smaller compared to the CH number density expected from LTE.

\begin{figure}
\centering
\includegraphics{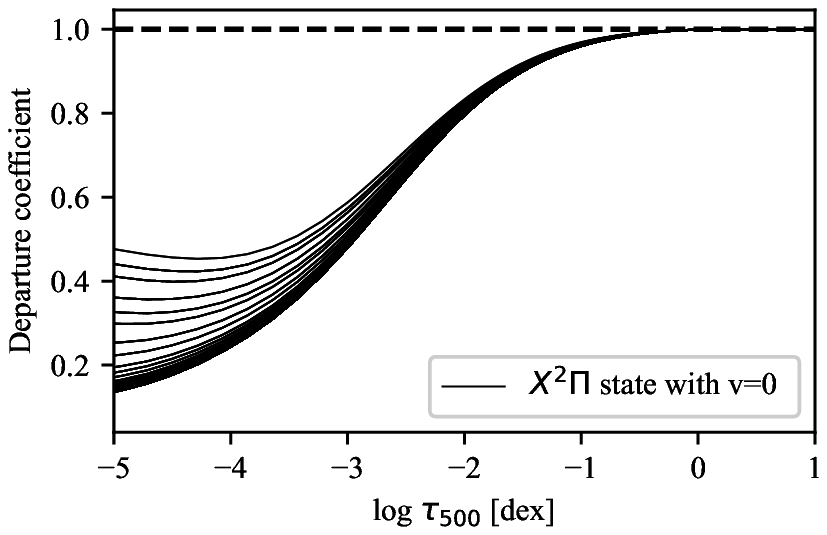}
\includegraphics{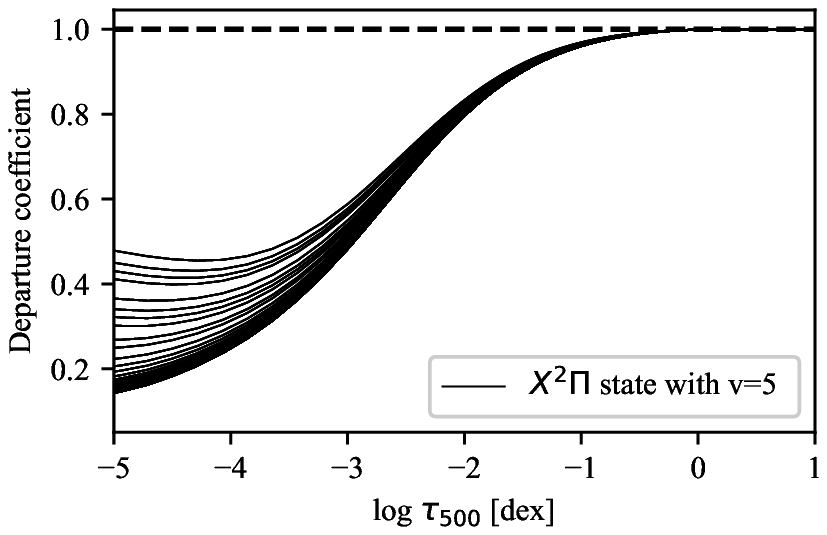}
\caption{Departure coefficients of levels in the ground state X\textsuperscript{2}$\Pi$ with vibrational quantum number $v=0$ (top) and $v=5$ (bottom), respectively. }
\label{fig:b_x}
\end{figure}

We have also investigated the departure coefficients of the level populations within a given electronic state of CH for the different rotational and vibrational levels. As shown in Fig. \ref{fig:b_x}, these energy states are not in complete thermal equilibrium. Even within the $\nu=0$ and $\nu=5$ vibrational  branches of the X\textsuperscript{2}$\Pi$ state the departure coefficients of the individual rotational levels slightly deviate from each other in the outer layers, at $\log \tau_{500} \lesssim -2.5$. At the same time, as also seen in Fig. 6, we support the finding by \citet{Hinkle1975} that the source functions of the transitions \textit{within} the ground state are thermal. This is because the populations of these states are primarily set by photo-dissociative processes, and the corresponding cross-sections are very similar.

It is not only important whether a level experiences departures from LTE or not, but also how the population of the transition's upper level relates to the population of the transition's lower level. To understand how exactly this  ratio influences the absorption line, it is useful to write the source function in terms of the departure coefficients $b_i$ (lower level) and $b_j$ (upper level) as follows:
\begin{equation}
    \frac{S_{\nu_0}}{B_\nu}\approx\frac{b_j}{b_i}\,,
    \label{eq:S/B}
\end{equation}
where $B_\nu$ is the Planck function.

\begin{table*}[t]
\begin{center}
\caption{NLTE abundance corrections for stellar model atmospheres with different metallicities [Fe/H] and carbon-to-iron abundance ratios [C/Fe].} 
\label{tab:correction}
\begin{tabular}{|c|c|c|c|}
\hline
 $T_\textnormal{eff}$ [K]/$\log g$/[Fe/H]& [C/Fe]& A(C)\textsubscript{NLTE} & $\Delta_{\textnormal{NLTE}}$ [dex] \\ 
\hline
4500/2.0/~~0.0  & 0.0 & 8.43  & $0.121 \pm 0.019$ \\
4500/2.0/$-$1.0 & 0.0 & 7.43 & $0.128 \pm 0.014$ \\
4500/2.0/$-$2.0 & 0.0 & 6.43  & $0.149 \pm 0.012$ \\
 & & & \\
4500/2.0/$-$3.0 & 0.0 & 5.43  & $0.180 \pm 0.016$ \\
4500/2.0/$-$3.0 & 0.7 & 6.13  & $0.167 \pm 0.015$ \\
4500/2.0/$-$3.0 & 1.5 & 6.93  & $0.175 \pm 0.042$ \\
4500/2.0/$-$3.0 & 3.0 & 8.43 & $0.179 \pm 0.007$ \\
 & & &  \\
4500/2.0/$-$4.0 & 0.0 & 4.43 & $0.210 \pm 0.024$ \\
4500/2.0/$-$4.0 & 0.7 & 5.13  & $0.192 \pm 0.018$ \\
4500/2.0/$-$4.0 & 1.5 & 5.93 & $0.185 \pm 0.020$ \\
4500/2.0/$-$4.0 & 3.0 & 7.43 & $0.148 \pm 0.013$ \\
\hline
\end{tabular}
\end{center}

\end{table*}

In the particular case of the representative CH line at 4299.48 {\AA}, in the deeper layers of the atmosphere both levels are thermalised, meaning that $b_j/b_i=1$, which leads to $S_{\nu_0}\approx B_\nu$ according to Eq. \ref{eq:S/B}. In other words, the source function is given by the Planck function, as it is the case under the LTE assumption. However, in the outer layers of the atmosphere both levels are 
under-populated ($b_i,\, b_j <0$), and the upper level has a higher departure coefficient than the lower level for all optical depths lower than $\log\tau_{500}\approx -1.5$. According to Eq. \ref{eq:S/B}, because $b_j>b_i$ and thus $b_j/b_i>1$, $S_{\nu_0}> B_\nu$ holds true in the upper photospheric layers. In other words, the source function exceeds the Planck function and is therefore super-thermal. A super-thermal source function is a consequence of the fact that there are more molecules in the excited upper level than in the lower level relative to the LTE distribution. This then leads to more emission than absorption compared to what would be expected under the assumption of LTE, leading also to the weaker NLTE absorption line than the LTE one in Fig. \ref{fig:line_0}.

Even though the molecular lines from the analysed region of the $G$-band are formed at slightly different depths in the atmosphere, and the molecular levels have varying departure coefficients, these coefficients follow the same pattern as those of the representative line displayed in Fig. \ref{fig:dep_0}, which implies that the upper levels have a higher departure coefficient than the lower levels. This then leads to more emission than would be expected from the LTE assumption, and the entire NLTE $G$-band is therefore weaker than the LTE $G$-band.

The fact that the LTE and NLTE profiles differ is expected to also influence the determined C abundances. The stronger LTE absorption line indicates that, in order for the LTE profile to match the NLTE profile, the C abundance has to be decreased in the LTE computation. This means that the carbon abundance is under-estimated when determined from the LTE line profile. Typically, the NLTE abundance correction is used to quantify the differences between LTE and NLTE line formation in terms of elemental abundance \citep{bergemannnordlander}:
\begin{equation}
\Delta_{\textnormal{NLTE}}=A(\textnormal{C})_{\textnormal{NLTE}}-A(\textnormal{C})_{\textnormal{LTE}}\,,
\end{equation}
where $A(\textnormal{C})_{\textnormal{NLTE}}$ and $A(\textnormal{C})_{\textnormal{LTE}}$ are the carbon abundances determined using the NLTE and the LTE computation, respectively. We obtain the NLTE correction  by varying the C abundance in the LTE calculation until the profile of the LTE line fits the line from the NLTE calculation. This is done for all individual lines in the $G$-band wavelength range of \makebox{4297 - 4303 {\AA}}. 
For the solar model, the averaged NLTE correction is then:
\begin{equation}
    \Delta_{\textnormal{NLTE}}=+0.038\pm0.001 \textnormal{~dex}
\end{equation}
This is a non-negligible difference given the standard requirements on the accuracy of solar abundances \citep[e.g.][]{Magg2022} and implies the NLTE effects should be taken into account in the analysis of the solar chemical composition.
\subsection{Metal-poor stars}
\begin{figure}
\centering
\includegraphics{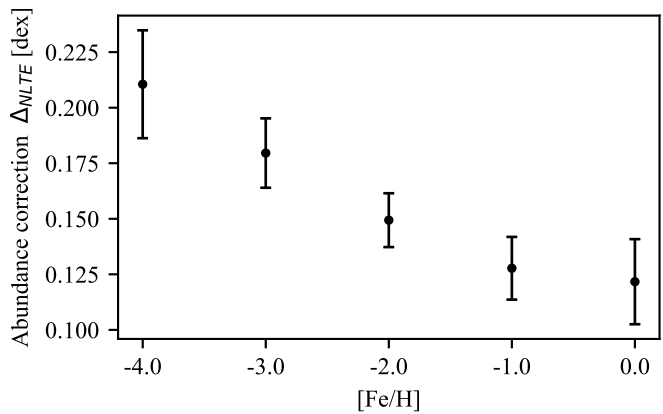}
\caption{NLTE abundance corrections for model atmospheres with \makebox{$T_\textnormal{eff}=4500$ K} and $\log g$ =2.0 with different metallicities, [Fe/H].}
\label{fig:corrections}
\end{figure}
The spectral features of the CH molecule are usually used for carbon abundance determinations in metal-poor stars \citep[e.g.][]{Aoki2007,CJHansen2016, Yoon_2016, spite2013}. 
In the following, we discuss the NLTE corrections computed using the model atmospheres described in Sect. 3.4. The results are presented in Table \ref{tab:correction}. For each such estimate, we also provide a line-to-line scatter that represents the differences in NLTE corrections for individual spectral lines within the $G$-band. 

The variation of NLTE abundance corrections with metallicity is shown in Fig. \ref{fig:corrections}. As emphasised earlier, here we focus on the parameters representative of the atmospheres of red giants, as many of them at low metallicity exhibit strong $G$-bands that are used as diagnostics of C abundances. Here, the [C/Fe] ratio is assumed to be scaled-solar, but we explore the dependence of NLTE effects on the assumed C abundance in the next section. Figure \ref{fig:corrections} clearly shows that the NLTE abundance corrections increase with decreasing metallicity. For the model atmosphere with the solar metallicity, the NLTE corrections are on average around $\Delta_\textnormal{NLTE} = +0.12 \pm 0.02$ dex. However, the NLTE correction increases to around $+0.2$ dex at [Fe/H]$=-4.0$. 
\begin{figure}
    \centering
    \includegraphics{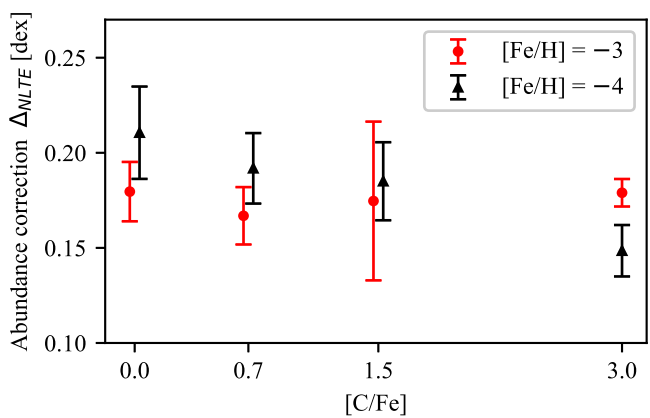}
    \caption{NLTE abundance corrections for metal-poor carbon-enhanced model atmospheres with metallicities  $\textnormal{[Fe/H]}=-3.0$ and $\textnormal{[Fe/H]}= -4.0$ and carbon-abundance ratios $\textnormal{[C/Fe] = 0.0, +0.7, +1.5, and +3.0}$.}
    \label{fig:corrections_CEMP_m}
\end{figure}
\begin{figure*}
    \centering
    \includegraphics{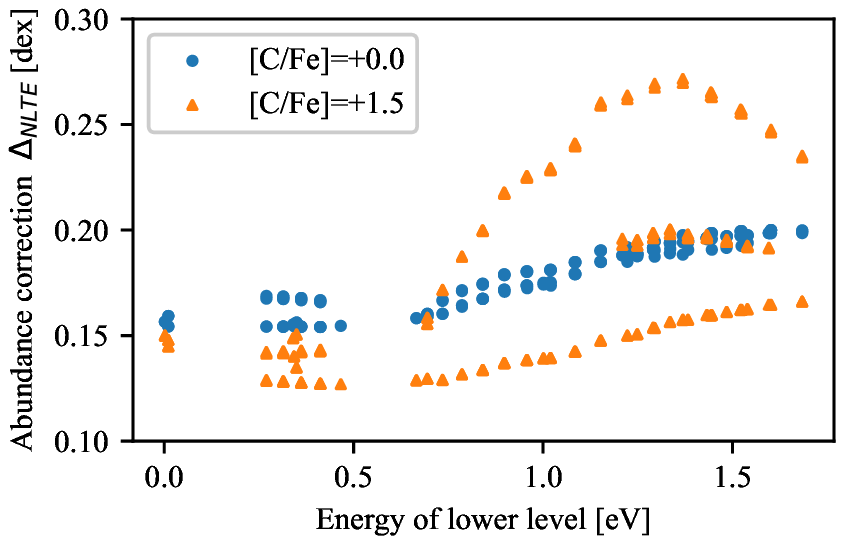}
    \includegraphics{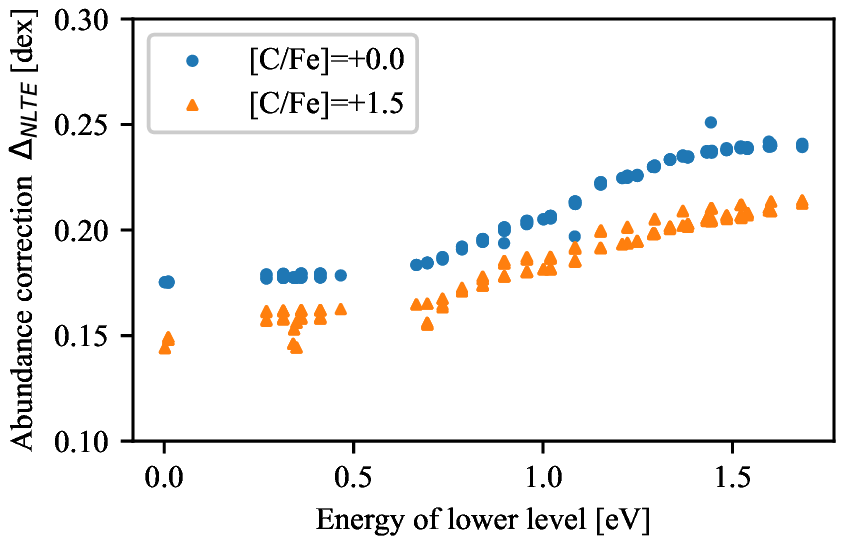}
    \caption{NLTE abundance corrections for individual $G$-band lines for model atmospheres with metallicity $\textnormal{[Fe/H]}=-3.0$ (left) and $\textnormal{[Fe/H]}=-4.0$ (right) for carbon-to-iron ratios $\textnormal{[C/Fe]}=+0.0, +1.5$. }
    \label{fig:corrections_CEMP}
\end{figure*}

Our results thus suggest that standard LTE analyses of the CH molecular lines in spectra of RGB stars under-estimate the abundance of C, and the bias is inversely proportional to metallicity, in the sense that NLTE effects are larger and more positive for lower-metallicity red giants.
\subsection{Carbon-enhanced metal-poor stars}
The NLTE analysis of the CH molecule is particularly important for C-enhanced metal-poor (CEMP) stars (Sect. 1). Especially at low metallicities, the CH molecule exhibits some of the strongest detectable features in the spectra.

The NLTE abundance corrections of the C-enhanced models are displayed in Fig. \ref{fig:corrections_CEMP_m} and also provided in Table \ref{tab:correction}. Here we focus on red giants with metallicities $\textnormal{[Fe/H]}=-3.0$ and $\textnormal{[Fe/H]}=-4.0$ and different [C/Fe] ratios ranging from 0 to 3 dex. Generally, the NLTE corrections are not very sensitive to the assumed C abundance. For the most metal-poor model,  $\textnormal{[Fe/H]}=-4.0$, the NLTE corrections decrease in amplitude with increasing [C/Fe] ratio. For the model atmosphere with [Fe/H]$=-3$, the NLTE $G$-band correction remains at the level of $+0.17$ dex for all [C/Fe] ratios considered. This implies that some non-carbon-enhanced metal-poor stars may in fact have higher C abundances than determined with LTE models.

Additionally, Fig. \ref{fig:corrections_CEMP} shows the NLTE abundance corrections for  all individual transitions in the wavelength range 4297 - 4303 {\AA}, as a function of the lower level energy, for the models that showed the largest variation in the corrections ($\textnormal{[C/Fe]}=+0.0, +1.5$ dex). For both metallicities the NLTE corrections increase for the transitions with a higher excitation potential, reaching $0.27$ dex for the C-enhanced ($\textnormal{[C/Fe]}=+1.5$) model atmosphere with metallicity $\textnormal{[Fe/H]}=-3$. The strong dependence on the NLTE corrections on the energy of the lower level may indicate that the temperature structure inferred from the excitation balance of CH lines may be prone to systematic biases.

\section{Discussion}
\label{sec:discussion}

We show that the $G$-band of the CH molecule experiences non-negligible NLTE effects for all the considered model atmospheres of K-type red giant stars apart from the Sun, with the NLTE effects increasing with decreasing metallicity of the model. This finding is important for all previous and future studies that use the spectrum of the CH molecule for 
carbon-abundance determinations in stellar atmospheres.

Based on these results, when using the spectral features of the CH molecule, the computed LTE carbon abundances for stars with stellar parameters similar to those considered in this work have to be revised, and adjusted by the corresponding NLTE corrections. For example, the determined LTE carbon abundance of a star with $T_\textnormal{eff} = 4500$ K, $\log g = 2.0$, $\textnormal{[Fe/H]} = -3.0$ and $\textnormal{[C/Fe]} = 0.0$ would have to be increased by 0.18 dex, which corresponds to a factor of 1.51.

As this is the first attempt at computing the $G$-band of the CH molecule in NLTE, no direct comparison with the literature can be made. The NLTE modelling of molecular bands is a quite recent field of research, and has not been attempted by many, mainly due to the complexities involved in modelling the spectrum of molecules due to the large number of energy levels and lines. An early NLTE analysis was done by \cite{CO_NLTE} for the $\Delta v=1$ band of the CO molecule in late-type atmospheres. Their model molecule is represented by 10 bound vibrational states each split in 121 rotational sub-states, which are connected by 1\,000 radiative transitions. Their results yield very small NLTE effects for the Sun, but larger effects for Arcturus. A NLTE analysis of the H\textsubscript{2}O molecule was carried out by \cite{Lambert} in the model atmospheres of red supergiants (RSG). They reported negative NLTE corrections for the H\textsubscript{2}O lines in the far infra-red ($\sim 3 \mu$m) spectrum of the RSG model. The inversion of NLTE effects is not 
unexpected, because the effects in such lines forming in transitions between high-excitation states are driven by recombination cascades and line scattering. Also other species, such as Si, Fe, Mg, and Ti, show positive NLTE effects in the optical spectra of FGK-type stars, but negative NLTE effects in the spectra of red supergiants 
\citep{bergemann2012c,Bergemann2013,Bergemann2015}.

Whereas our NLTE analysis is carried out with 1D hydrostatic model atmospheres, it should be pointed that effects of 3D convection may also be important in the diagnostic spectroscopy based on molecular lines. A detailed LTE analysis of the $G$-band with 1D and 3D models of red giant was carried out by \cite{Collet2007}. They found that the cooler outermost structures of the 3D models favour a higher concentration of CH molecules, which in turn leads to stronger $G$-band features and consequently to lower C abundances inferred in 3D, compared to standard modelling with 1D LTE models (see their Table 3). A similar result was reported by \citet{CH3D} for the synthetic $G$-band in metal-poor dwarfs, where Table 2 of their paper shows that the majority of the 3D corrections are negative. This means that in LTE line formation 1D hydrostatic models over-estimate the carbon abundances.

Our results suggest that the NLTE effects in the CH molecule are driven by over-dissociation, owing to the low photo-dissociation threshold of the molecule, and thus the effects will increase with decreasing metallicity and lower density of the model atmospheres. It is  therefore not yet clear whether 3D and NLTE effects will compensate for each other, or if an amplification of the effect would result. We therefore \textit{caution against co-adding 3D and NLTE corrections, as is sometimes done in the literature}. The LTE 1D assumption may indeed predict the correct C abundance, but for the wrong reasons. A detailed quantitative comparison between the 3D and NLTE effects, ideally including full 3D NLTE radiative transfer calculations, is necessary in order to understand which of the effects would dominate the formation of the $G$-band.

\section{Conclusions}
\label{sec:conclusion}
In all the previous studies of the CH molecule in the atmospheres of FGK-type stars that we are aware of, the theoretical CH spectral lines were computed using the assumption of LTE. This work presents the first NLTE radiative transfer calculation for the CH molecule under the conditions of metal-poor and carbon-enhanced stellar atmospheres. Our primary aim is to explore the difference between the LTE and the NLTE computation. We focus on the optical $G$-band of the CH molecule, the feature commonly used for C-abundance determinations, especially for cool metal-poor stars. 

Our new NLTE model of the CH molecule includes 1981 levels and 53079 radiative transitions adopted from \cite{Masseron}. The photo-dissociation cross-sections are taken from \cite{photodisskurucz}, and collisional reactions are represented by classical semi-empirical recipes such as Regemorter's \citep{Regemorter} and Allen's  formula \citep{Allen} for collisions with electrons, and Drawin's formula \citep{Steenbock} for collisions with H atoms. The detailed statistical equilibrium calculations were performed using the MULTI2.3 code and 1D LTE MARCS model atmospheres. 

We find that the $G$-band NLTE line profiles are systematically weaker compared to their LTE counterparts, regardless of the stellar model atmosphere used. This has a systematic effect on the C-abundance determination. The LTE assumption leads to 
\textit{under-estimated} C abundances for all considered model atmospheres, by $\approx +0.1$ dex for solar-metallicity giants and $\approx +0.2$ dex for metal-poor red giants with [Fe/H] $\lesssim -3.0$. Increasing the [C/Fe] abundance ratio of the model leads to slightly smaller, by $\sim 0.06$ dex, NLTE abundance corrections, which may be relevant to studies of CEMP stars.

We also find that, even though the $G$-band averaged NLTE corrections are not that large, when inspecting the individual NLTE corrections for the C-enhanced models in Fig. \ref{fig:corrections_CEMP_m}, we can see that for some particular lines, especially for those arising from transitions with higher energy of the lower level, the NLTE corrections can exceed $+0.25$ dex. This implies the C abundance would be 
under-estimated by a factor of 1.8.

The results of this work show that NLTE effects are non-negligible for all considered stellar model atmospheres, suggesting that LTE is not a reliable assumption for the modelling of the CH molecular bands in spectra of cool FGK-type stars.

We note, however, that some important molecular data are still missing. In particular, the collision-induced reactions with electrons and hydrogen atoms require quantum-mechanical treatment, and photo-dissociation cross-sections for higher energy states of the CH molecule are needed. 
It is also important to mention that we study NLTE effects in the CH molecule in isolation, whereas in reality there are other competing molecules, like CO, that might affect its abundance and behaviour. Studies of non-equilibrium chemistry with NLTE effects are, however, beyond the scope of the present investigation.

We intend to expand our study to include model atmospheres with larger ranges of effective temperatures and surface gravities, so that a more comprehensive set of NLTE corrections can be derived for application to the derivation of [C/Fe] for CEMP stars in the future.

\begin{acknowledgements}
T.C.B. acknowledges partial support for this work from grant PHY 14-30152; Physics Frontier Center/JINA Center for the Evolution of the Elements (JINA-CEE), and OISE-1927130: The International Research Network for Nuclear Astrophysics (IReNA), awarded by the US National Science Foundation. MB is supported through the Lise Meitner grant from the Max Planck Society. We acknowledge support by the Collaborative Research centre SFB 881 (projects A5, A10), Heidelberg University, of the Deutsche Forschungsgemeinschaft (DFG, German Research Foundation). This project has received funding from the European Research Council (ERC) under the European Union’s Horizon 2020 research and innovation programme (Grant agreement No. 949173).
\end{acknowledgements}

\bibliographystyle{aa} 
\bibliography{references} 

%
%

\end{document}